\documentclass[
twocolumn,
pra,showpacs,preprintnumbers,amsmath,amssymb]{revtex4}

\usepackage{times}
\usepackage{graphicx}
\usepackage{amssymb}
\usepackage{amsthm}
\usepackage{amsmath}
\usepackage{dsfont}
\usepackage{bm}
\usepackage{mathrsfs}
\usepackage{bbold}

\begin{document}

\title{Cherenkov Radiation of Light Bullets}
\author{Ulf Leonhardt and Yuval Rosenberg}
\affiliation{
\normalsize{
Department of Physics of Complex Systems,
Weizmann Institute of Science, Rehovot 761001, Israel}
}
\date{\today}

\begin{abstract}
Electrically charged particles, moving faster than the speed of light in a medium, emit Cherenkov radiation. Theory predicts electric and magnetic dipoles to radiate as well, with a puzzling behavior for magnetic dipoles pointing in transversal direction [I. M. Frank, Izv. Akad. Nauk SSSR, Ser. Fiz. {\bf 6}, 3 (1942)]. A discontinuous Cherenkov spectrum should appear at threshold, where the particle velocity matches the phase velocity of light. Here we deduce theoretically that light bullets [Y. Silberberg, Opt. Lett. {\bf 15}, 1282 (1990)] emit an analogous radiation with exactly the same spectral discontinuity for point--like sources. For extended sources the discontinuity turns into a spectral peak at threshold. We argue that this Cherenkov radiation has been experimentally observed in the first attempt to measure Hawking radiation in optics [F. Belgiorno {\it et al.}, Phys. Rev. Lett. {\bf 105}, 203901 (2010)] thus giving experimental evidence for a puzzle in Cherenkov radiation instead.

\end{abstract}

\maketitle

\section{Introduction}

\begin{figure}[b]
\begin{center}
\includegraphics[width=20.pc]{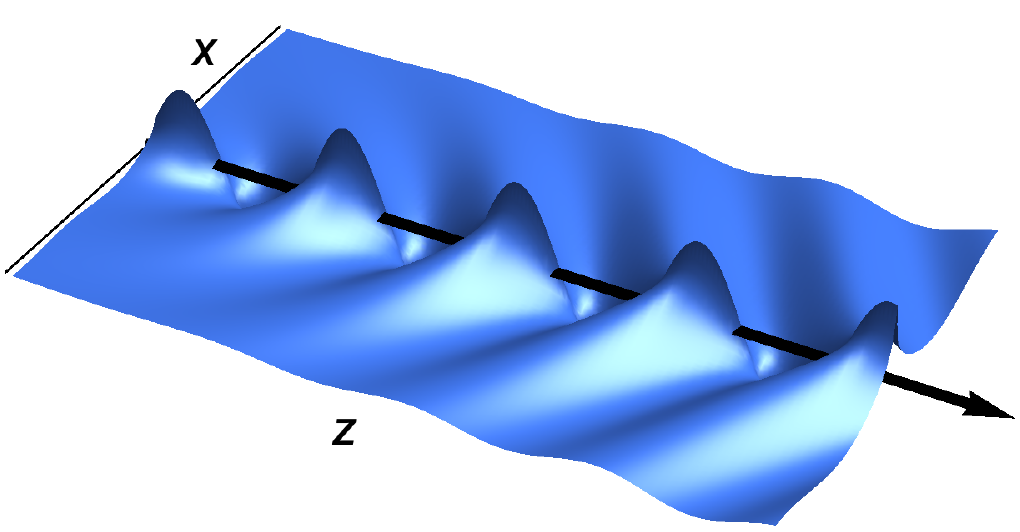}
\caption{
\small{
Cherenkov radiation emitted from a light bullet traveling in $z$--direction. The figure shows one Fourier component of the radiation field in the $x$,$z$ plane (units in $\mu\mathrm{m}$). The real part of $\widetilde{A}_x$ is plotted with $\widetilde{A}_x$ described in Fig.~\ref{phase}. The field appears to be made by a Fourier--transformed light bullet, as if the bullet were a train of sources oscillating with a fixed frequency. The figure illustrates the Mach--cone emission of light in analogy to the supersonic boom and also to the bow wave of ships. 
}
\label{field}}
\end{center}
\end{figure}

Electrically charged particles radiate when they move faster than the phase velocity of light in a medium. Cherenkov \cite{Cherenkov} was the first to experimentally investigate this radiation starting in 1934 (thanks to his excellent eye sight). These days, Cherenkov radiation is clearly visible in the eerie blue light in nuclear reactors \cite{Reactors}, and is widely known and used, from research in high--energy and astrophysics \cite{ReviewCherenkov,IceCube} to medical applications \cite{ReviewMed}. Frank and Tamm \cite{FT} realized that Cherenkov radiation is a consequence of classical electromagnetism in media \cite{LL8} (after similar ideas by Heaviside \cite{Heaviside} and Sommerfeld \cite{Sommerfeld} were forgotten) and determined theoretically the emitted spectrum, which agreed with refined experimental measurements. Figure \ref{field} shows the emitted radiation field (for the case considered in this paper). The figure illustrates how closely Cherenkov radiation is related to the physics of the supersonic boom \cite{LL6} and the bow wave of ships \cite{CambridgeFluids}. 

\begin{figure}[t]
\begin{center}
\includegraphics[width=19.pc]{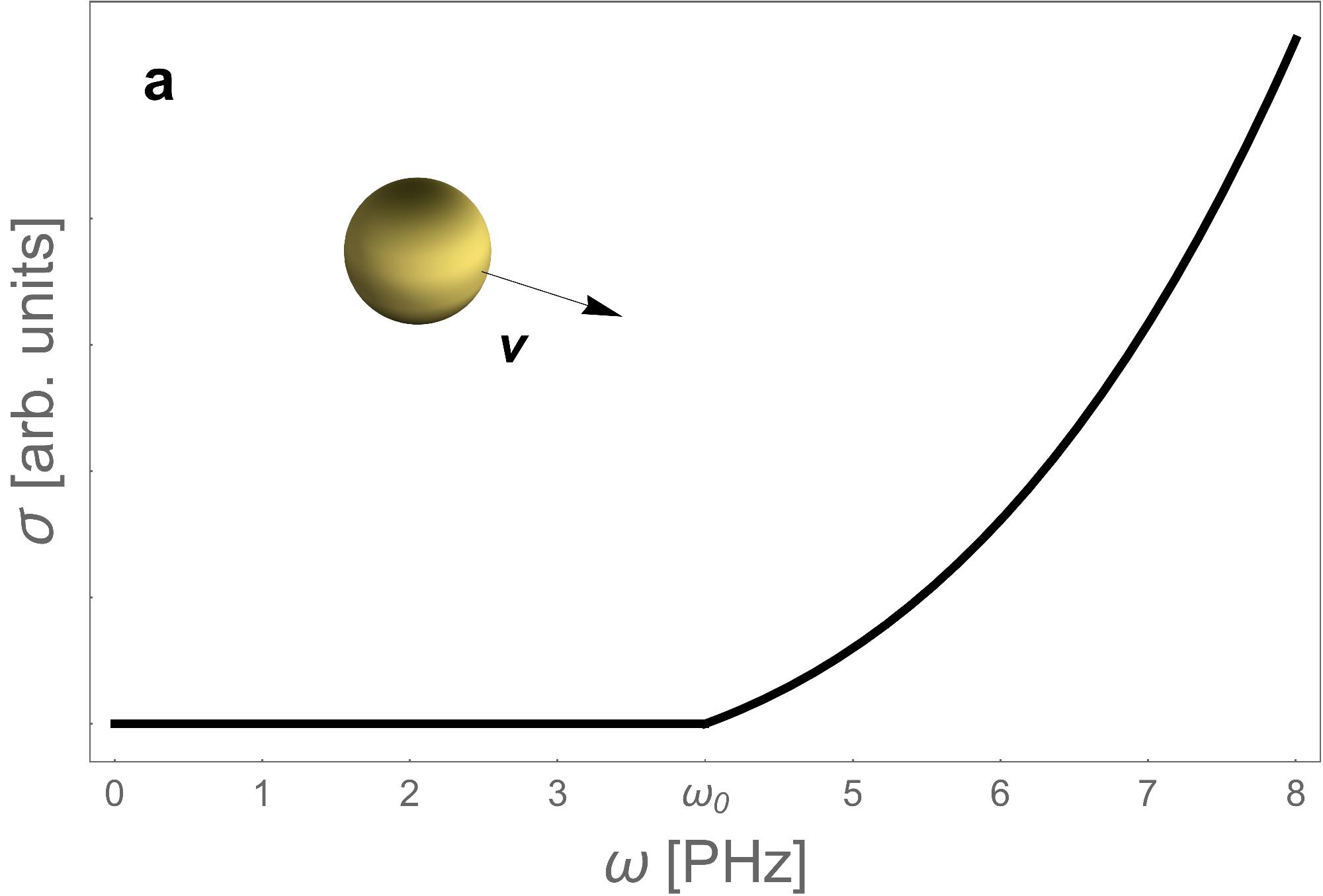}
\includegraphics[width=19.pc]{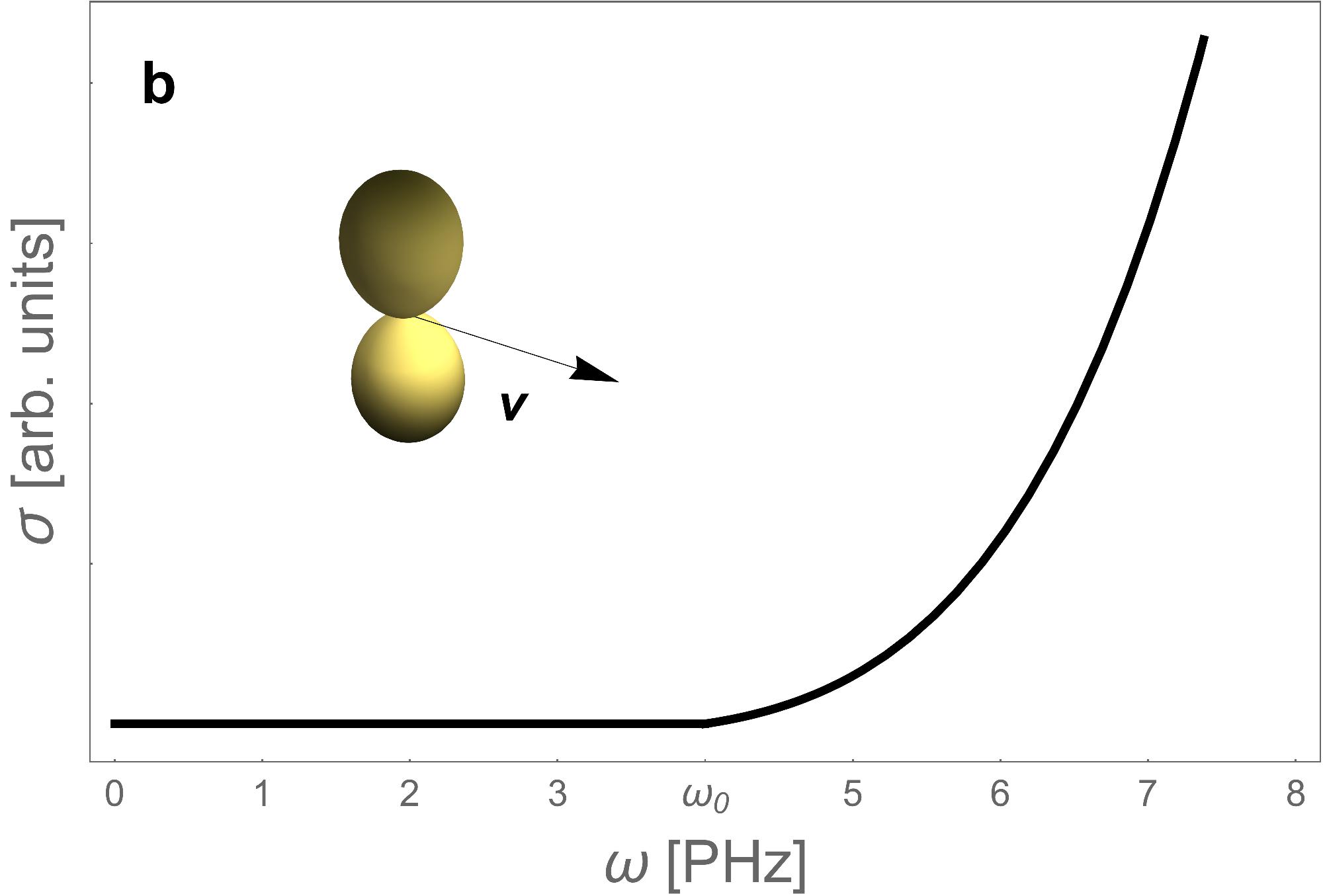}
\includegraphics[width=19.pc]{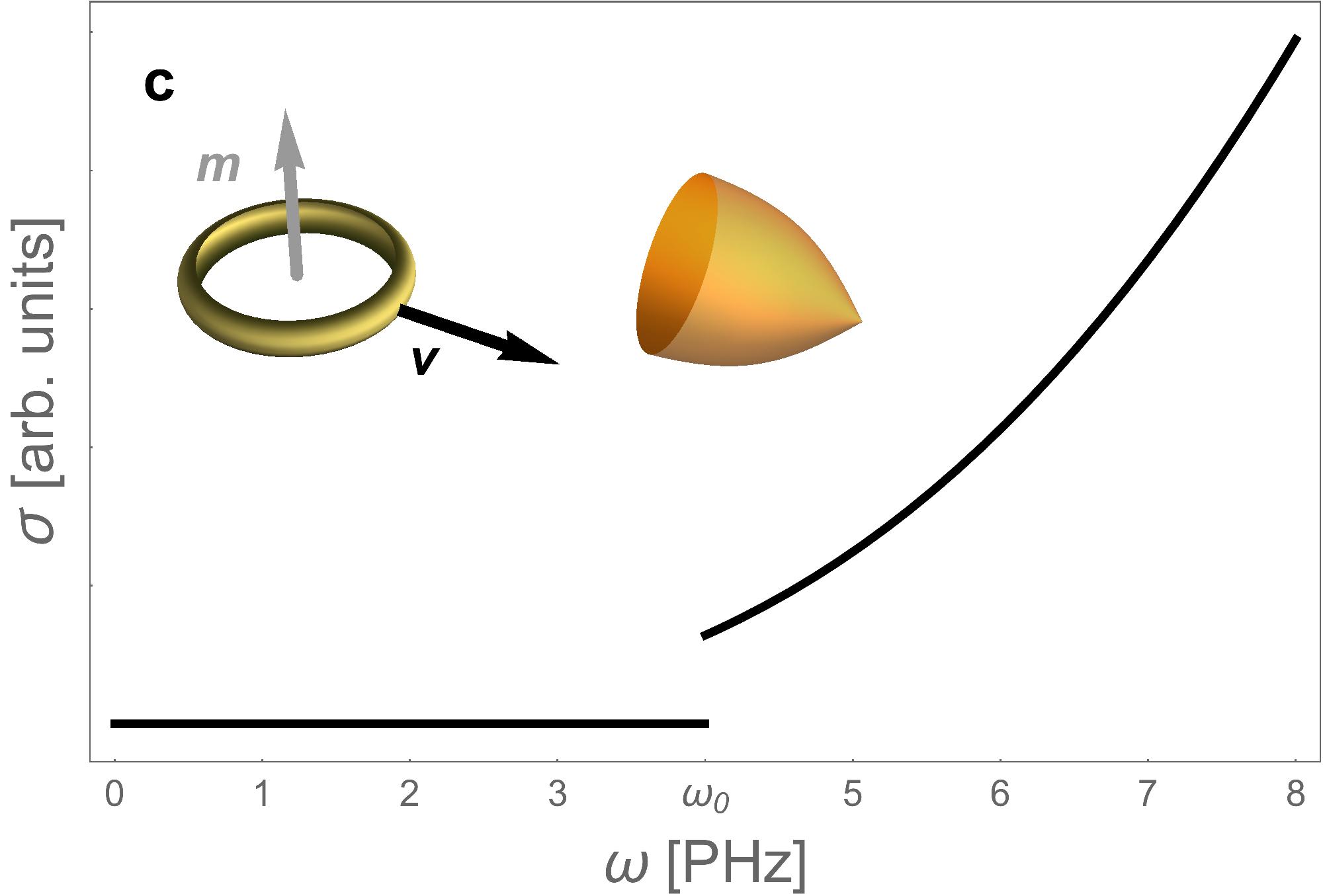}
\caption{
\small{
Cherenkov spectra. Particles are moving in glass with a velocity $v$ that matches the phase velocity at frequency $\omega_0=4.0 \mathrm{PHz}$. The refractive index is given by the standard Sellmeier formula for fused silica \cite{Sellmeier}. The figures show the emission spectra $\sigma$ [Eq.~(\ref{sigma})] for various types of particles. {\bf a}: electric monopoles, {\bf b}: electric dipoles, {\bf c}: magnetic dipoles pointing in transversal direction represented by the light bullets of this paper. In the case {\bf{c}} the emission spectrum jumps at the onset of Cherenkov radiation. 
}
\label{spectra}}
\end{center}
\end{figure}

It is natural to generalize the Cherenkov radiation of charged particles --- monopoles --- to the radiation of dipoles or higher--order multipoles. Frank  \cite{Frank} predicted the Cherenkov spectrum of electric and magnetic dipoles, and found a surprize: while the Cherenkov spectrum of electric monopoles and dipoles rises smoothly from zero when the particle becomes superluminal (Figs.~\ref{spectra}a and \ref{spectra}b), the spectrum of magnetic dipoles polarized orthogonally to the direction of motion, suddenly jumps from zero (Fig.~\ref{spectra}c). In all cases shown (Fig.~\ref{spectra}), the spectra depend on the refractive index $n$ of the host medium and the velocity $v$ of the particle. As $n$ depends on the frequency $\omega$ of light, the onset of Cherenkov radiation occurs at the critical frequency where $c/n$ matches $v$. The discontinuous onset for magnetic dipoles seemed unphysical and puzzled Frank throughout his life \cite{Frankreview}. Since there has been no experimental evidence for the Cherenkov radiation of magnetic dipoles, this puzzle remained without full resolution.

In this paper, we consider a closely related case of Frank's fast--moving magnetic dipoles that occurs in nonlinear optics \cite{NLO} and that, as we will explain, has probably been observed experimentally. It has been known that Cherenkov radiation can also be formed by electromagnetic radiation in media, rather than by moving particles. This idea was first realized by Askar'yan \cite{Askar'yan} who analyzed the induced material polarization at the electric field gradients. Later, the idea was further developed and experimentally verified in electro--optic materials \cite{EO_Theory,EO_exp}, that posses DC electric polarization and create Cherenkov--type emission similar to that of relativistic electric dipoles. Many other phenomena were also related to the Cherenkov effect, while some possess only a Cherenkov--type phase matching, without the shock--wave nature of Cherenkov emission (see e.g.\ Ref.~\cite{DW}). Note that this distinction is not always straightforward (see e.g.\ Ref.~\cite{Phase-matching_not-cherenkov}).

Here we consider optical pulses called light bullets \cite{Silberberg}. In nonlinear media the Kerr effect \cite{NLO} can hold light pulses together, forming light bullets \cite{Silberberg}. We show that the nonlinear electric polarization of such pulses in media acts like a combination of an electric and a magnetic dipole, both pointing in transversal direction. We determine theoretically the spectrum of a point--like source and find the same sudden onset (Fig.~\ref{spectra}c) as the one predicted by Frank. Moreover, we point out that this Cherenkov radiation of light bullets has probably been seen in the first heroic attempt \cite{Faccio1} to demonstrate Hawking radiation \cite{Hawking} in optics. There, instead of Hawking radiation, a distinct peak was seen in the spectrum where the phase velocity $c/n$ matched the speed $v$ of the light bullet \cite{Faccio1}. After initial discussion \cite{Schmutz} the peak was later attributed to some form of superluminal emission \cite{FaccioSupp}. Here we make this relationship more precise: we relate the peak to the sudden onset of Cherenkov radiation as the feature surviving optical interference. The connection to Cherenkov radiation has been dismissed before \cite{FaccioSupp}, but the point is this: unlike the idealized case of Frank's magnetic dipole, a real light bullet is an extended object. On a whole surface in the light bullet the Kerr effect has enhanced the refractive index such that it matches the bullet's velocity. The Cherenkov radiation emitted at various positions inside the light bullet interferes. Using a simple model for a finite emission disk we find that the onset of Cherenkov radiation is the only surviving feature of the radiation, turning Frank's jump in the spectrum into a peak --- probably the observed one \cite{Faccio1}.

\section{The model}

Let us first focus on the idealized model of the light bullet as a moving point object. The oscillating electric field of the bullet creates in the host medium an oscillating electric polarization pointing in transversal direction. The linear part of this dielectric polarization gives rise to the linear electric susceptibility and hence the deviation of the refractive index $n$ from unity. The nonlinear part may be viewed \cite{Philbin} as generating a perturbation of the refractive index that is proportional to the pulse intensity and hence moves with the pulse. The dielectric polarization of this perturbation is the moving object whose radiation we are going to investigate here. 

Consider a point object of nonlinear electric polarization $\bm{P}$  pointing in $x$--direction and moving with velocity $v$ in $z$--direction (Fig.~\ref{field}). Changes in $\bm{P}$ generate electric currents \cite{Jackson} with density
\begin{equation}
{\bm j} = \partial_t {\bm P}
\label{j}
\end{equation}
with
\begin{equation}
P_x= P\, \Theta(z)\,\delta(z_0)\,\delta(x)\,\delta(y) \,,\quad P_y=P_z=0 
\label{p}
\end{equation}
at the moving position
\begin{equation}
z_0=z-vt \,.
\label{zprime}
\end{equation}
We assume from the outset that $v$ exceeds the phase velocity $c/n$ in the medium, for otherwise no Cherenkov radiation can be generated on general grounds \cite{Condition}. The polarization is switched on \cite{TransitionRad} at some position where the pulse enters the medium, here $z=0$, as the Heaviside function $\Theta(z)$ indicates. We assume that the subsequently produced radiation may propagate in an infinite medium, ignoring surface effects, for keeping the calculations simple (but not too simple). 

Let us compare this model with Frank's moving dipoles \cite{Frankreview}. These dipoles are electric or magnetic dipoles in their co--moving frame. For comparing them with our case, we thus need to transform the charge and current profiles of the moving polarization [Eq.~(\ref{p}) and (\ref{zprime})] to its co--moving frame. In relativity \cite{LL2} the charge and current densities $\varrho$ and $\bm{j}$ form the density of the contravariant four--vector $(c\varrho, \bm{j})$ where, in our case, 
\begin{equation}
\varrho = - \nabla\cdot\bm{P} = -\partial_x P_x
\end{equation}
and $\bm{j}$ is given by Eq.~(\ref{j}). The Lorentz transformation \cite{LL2} to the co--moving frame modifies the charge density 
\begin{equation}
\varrho' = \gamma \varrho = -\partial_x\gamma P_x
\end{equation}
by the relativistic gamma factor $\gamma =(1-v^2/c^2)^{-1/2}$ but does not introduce a contribution from the current, as $\bm{j}$ points orthogonally to the direction of motion. The $x$--component of the current density remains unaffected as well, but we need to transform the time derivative in Eq.~(\ref{j}) to the co--moving frame, and get
\begin{equation}
j_x' = -v\,\partial_z'\gamma P_x
\end{equation}
as $\partial_t'P_x=0$. The $z$--component of the current density receives, in the Lorentz transformation,  a contribution from the charge density such that
\begin{equation}
j_z' = -v\gamma\varrho = v\,\partial_x \gamma P_x \,.
\end{equation}
We thus obtain for the current density 
\begin{equation}
\bm{j}'=\nabla'\times\bm{M} \,,\quad M_y= v\gamma P_x
\end{equation}
and vanishing $M_x$ and $M_z$ components. This is the effective current density of a transverse magnetic dipole with magnetization $\bm{M}$ \cite{Jackson}. The charge density $\varrho'$, on the other hand, describes a transverse electric dipole. Our model thus corresponds to a certain combination of an electric with a magnetic dipole. As we will show in Sec.~V this combination produces a Cherenkov spectrum that is simpler than the spectrum of the magnetic dipole alone. We are also going to reproduce the discontinuity at the threshold of Cherenkov radiation. As the spectrum of the moving electric dipole is continuous \cite{Frankreview}, our analysis supports Frank's discontinuity of the magnetic Cherenkov radiation \cite{Frankreview}.

Having established our model, we proceed to laying the ground for our calculations of the Cherenkov spectrum. The electric and magnetic field strengths ${\bm E}$ and ${\bm H}$ in SI units are given in terms of the electromagnetic potentials as
\begin{equation}
{\bm E} = -\partial_t {\bm A} - \nabla{U} \,,\quad \mu_0{\bm H} = \nabla\times{\bm A} 
\label{eh}
\end{equation}
with magnetic permeability $\mu_0$. It will be advantageous to impose the Lorentz gauge \cite{Jackson}:
\begin{equation}
\nabla\cdot {\bm A} + \frac{n^2}{c^2} \,\partial_t U = 0 \,.
\label{lorentz}
\end{equation}
From Maxwell's equations follows \cite{Jackson}
\begin{equation}
\nabla^2 {\bm A} - \frac{n^2}{c^2}\,\partial_t^2 {\bm A} = \mu_0 \partial_t {\bm P} \,,
\label{max}
\end{equation}
which implies, {\it inter alia}, that the vector potential ${\bm A}$ points in $x$--direction as well. 

In order to define a spectrum, we Fourier--transform the only non--trivial vector potential component:
\begin{equation}
\widetilde{A}_x=\int_{-\infty}^{+\infty} A_x \,e^{i\omega t} \, dt
\label{fourier}
\end{equation}
and obtain from Eqs.~(\ref{j}-\ref{zprime}), and Eqs.~(\ref{max}) and (\ref{fourier}):
\begin{equation}
\left(\nabla^2 + n^2k^2\right)\widetilde{A}_x = \frac{i\omega \mu_0 P}{v}\, e^{i k z} \delta(x)\,\delta(y) 
\label{ax}
\end{equation}
in terms of the free--space wavenumber defined as
\begin{equation}
k = \frac{\omega}{c} \,.
\end{equation}
Due to the cylindrical symmetry of Eq.~(\ref{ax}), $\widetilde{A}_x$ will only be a function of $z$ and $r$ with $r^2=x^2+y^2$. In order to fully take advantage of the cylindrical symmetry we use cylindrical coordinates $\{r,\phi,z\}$ with metric $dl^2=dr^2+r^2d\phi^2+dz^2$ and $g=r^2$ for the determinant of the metric tensor. We get from the Lorentz gauge, Eq.~(\ref{lorentz}), 
\begin{equation}
\widetilde{U} = - \frac{i c}{n^2k}\,\partial_x \widetilde{A}_x = - \frac{i c}{n^2k}\,\partial_r \widetilde{A}_x \cos\phi
\label{u}
\end{equation}
as $\partial r/\partial x = x/r = \cos\phi$. These first mathematical consequences from our simple model have prepared us for calculating the electromagnetic field. 

\section{Vector potential}

The main mathematical ingredient of the calculation is the scalar Green function $G$ describing the field of an instantaneous polarization, according to the propagation equation 
\begin{equation}
\left(\nabla^2 - \frac{n^2}{c^2}\,\partial_t^2\right)G = -\partial_t \delta(t)\,\delta(x)\,\delta(y)\,\delta(z) 
\label{green}
\end{equation}
with the well--known solution \cite{Jackson}
\begin{eqnarray}
\widetilde{G} &=& - \frac{i\omega}{4\pi\rho}\,e^{ink\rho} \,,
\label{gsolution} \\
\rho &=& \sqrt{r^2+z^2}
\label{rho}
\end{eqnarray}
for the Fourier--transformed, outgoing $\widetilde{G}$. Writing
\begin{equation}
\Theta(z)\,\delta(z-vt)= \int_0^\infty \delta(t-t_0)\,\delta(z-vt)\,dt_0
\end{equation}
we obtain from Eqs.~(\ref{j}-\ref{zprime}), (\ref{max}) and (\ref{green})
\begin{equation}
A_x = \mu_0 P \int_0^\infty G(z-vt_0,t-t_0) \,dt_0 \,,
\end{equation}
and hence
\begin{eqnarray}
\widetilde{A}_x &=& \mu_0 P \int_0^\infty \widetilde{G}(z-vt_0)\,e^{i\omega t_0}\,dt_0 
\nonumber\\
&=& \frac{\mu_0 P}{v}\, e^{i k z} \int_{-\infty}^{z} \widetilde{G}(z_0)\,e^{-i k z_0} dz_0 
\label{axg}
\end{eqnarray}
with $z_0=z-vt_0$. Inserting Eq.~(\ref{gsolution}) we get the explicit expression
\begin{eqnarray}
\widetilde{A}_x &=& -\frac{i\omega \mu_0 P}{4\pi v} \, \int_{-\infty}^z \exp\left(i\frac{\omega n \rho}{c} + i\frac{\omega}{v}\,(z-z_0)\right)\frac{dz_0}{\rho} 
\nonumber\\
&& \mbox{with}\quad \rho = \sqrt{r^2+z_0^2} \,.
\end{eqnarray}
We define the wave--number walk--off $\delta k$ as
\begin{equation}
\delta k \equiv \frac{\omega}{c}\sqrt{n^2-\frac{c^2}{v^2}}
\label{deltak}
\end{equation}
for superluminal propagation, $v>c/n$, as assumed, and represent the phase of the integrand as
\begin{equation}
\delta k\, r \cosh\chi \equiv \frac{\omega n \rho}{c} - \frac{\omega z_0}{v} 
\label{chi}
\end{equation}
where $\chi$ is a function of $z_0$ defined by this equation. Since 
\begin{eqnarray}
\delta k\, r \sinh\chi\,d\chi &=& \left(\frac{\omega n z_0}{c\rho} - \frac{\omega}{v} \right) dz_0 \,,
\nonumber\\
(\delta k)^2 r^2 \sinh^2\chi &=& \left(\frac{\omega n z_0}{c} - \frac{\omega \rho}{v}\right)^2
\end{eqnarray}
we find
\begin{equation}
d\chi = \frac{dz_0}{\rho}
\end{equation}
and therefore 
\begin{equation}
\widetilde{A}_x =  -\frac{i\omega \mu_0 P}{4\pi v}\, e^{i(\omega/v)z} \int_{-\infty}^{\chi_0} e^{i\delta k r \cosh\chi}\, d\chi 
\label{axx}
\end{equation}
where $\chi_0$ follows from Eqs.~(\ref{rho}), (\ref{deltak}) and (\ref{chi}) as
\begin{equation}
\cosh \chi_0 = \frac{nv\sqrt{r^2+z^2}-cz}{r\sqrt{n^2 v^2 -c^2}} \,.
\label{cosh}
\end{equation}
Using identities of hyperbolic functions we obtain
\begin{equation}
\chi_0 = \mathrm{artanh}\frac{z}{\sqrt{r^2+z^2}} - \mathrm{artanh}\frac{c}{nv} \,.
\label{chi0}
\end{equation}
Note that Eq.~(\ref{cosh}) has two real solutions, a positive and a negative one; we have chosen in Eq.~(\ref{chi0}) the branch with $\chi_0<0$ at $z=0$ for reasons that are going to become clear in the next paragraph. 

Having established the solution for the Fourier--transformed vector potential $\widetilde{A}_x$, we express it in a physically intuitive and numerically convenient form. We use the integral representation of the Hankel function \cite{Erdelyi}, 
\begin{equation}
H_0^{(1)}(\xi)=\frac{1}{i\pi} \int_{-\infty}^{+\infty} e^{i\xi \cosh\chi}\, d\chi 
\end{equation}
and obtain from Eq.~(\ref{axx}):
\begin{equation}
\widetilde{A}_x = \frac{\omega \mu_0 P}{4v} \, H\,e^{i(\omega/v)z}
\label{axh}
\end{equation}
with the definition
\begin{equation}
H \equiv H_0^{(1)}(\delta k\,r) - \frac{1}{i\pi} \int_{\chi_0}^\infty e^{i\delta k r \cosh\chi}\, d\chi \,.
\label{h}
\end{equation}
The Hankel function $H_0^{(1)}$ describes in Eq.~(\ref{axh}) a stationary, outgoing radiation emitted along the propagation axis of the source, the $z$--axis, because of the asymptotics \cite{Erdelyi}
\begin{equation}
H_0^{(1)}(\xi) \sim \sqrt{\frac{2}{\pi\xi}}\, e^{i(\xi-\pi/4)} \quad\mbox{for large $\xi$} \,.
\label{asymphankel}
\end{equation}
The remaining integral in Eq.~(\ref{h}) falls off for $z\rightarrow\infty$ where $\chi_0\rightarrow\infty$ according to Eq.~(\ref{chi0}). The integral thus contains the transition radiation \cite{TransitionRad} emitted upon the polarization has entered the medium at $z=0$. The saddle point of the integrand's phase lies at $\chi=0$, so only for $\chi_0<0$ the correcting integral will play a major role in the far field. Since near the entrance of the moving polarization the field needs to be strongly modified from stationary radiation, we have chosen the branch of $\chi_0$ in such a way that $\chi_0<0$ for $z=0$.  Note that the integral also describes corrections [Eq.~(\ref{hasymp})] due to near--field effects. 

\begin{figure}[t]
\begin{center}
\includegraphics[width=20.pc]{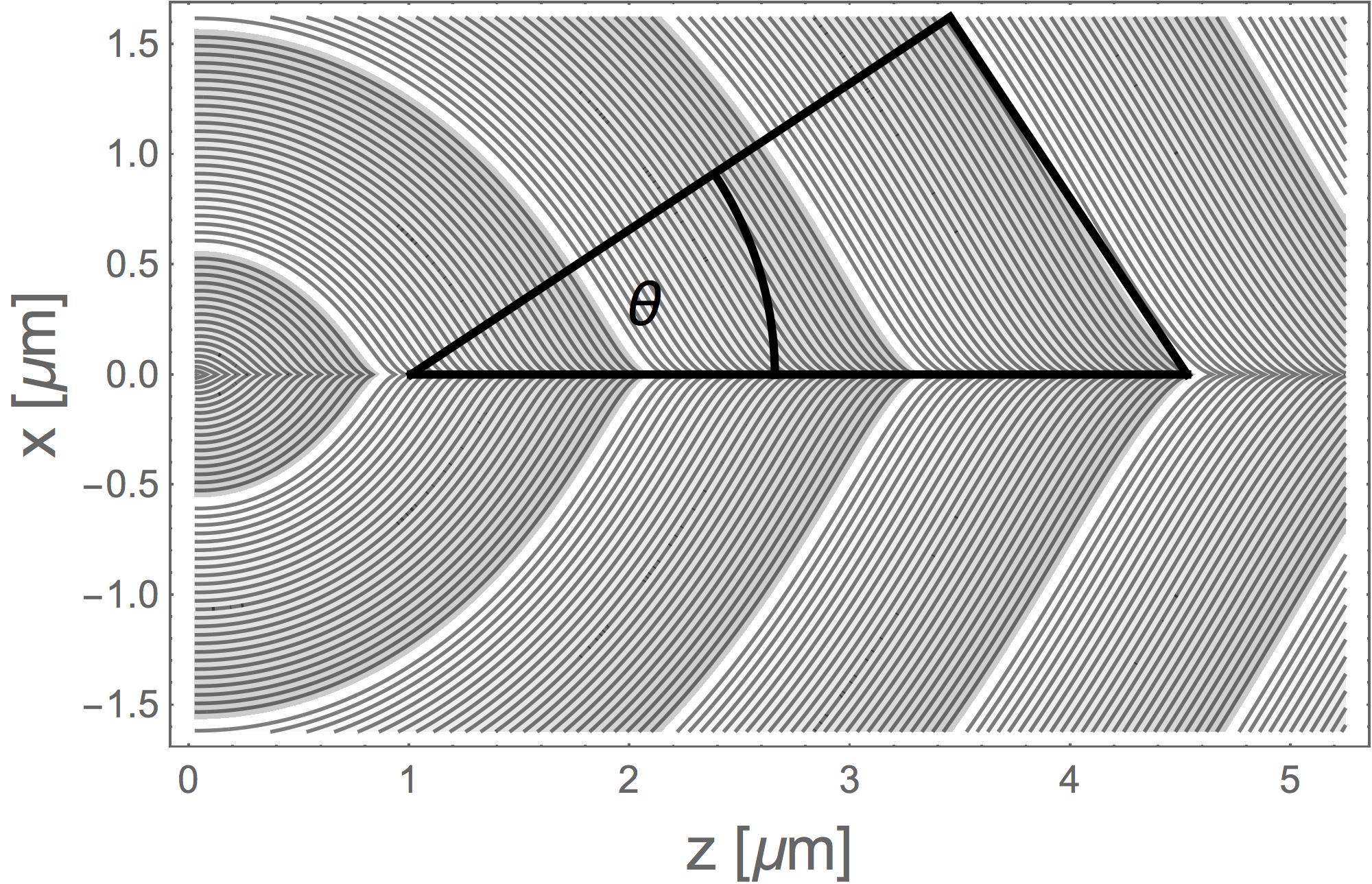}
\caption{
\small{
Phase pattern of Cherenkov radiation showing the Mach cones of the emitted radiation and their modification due to the near field and the entrance of the point source. Near the entrance ($z=0$) the cones morph into the spherical waves of transition radiation \cite{TransitionRad}. The plot shows the contour lines of the argument of $\widetilde{A}_x$ given by Eqs.~(\ref{axh}), (\ref{hnum}) and (\ref{deltak}) with $n=1.45$, $v=1.2c/n$, $\omega=2\pi c/\lambda$, $\lambda=1.5\mu\mathrm{m}$ (as in Fig.~\ref{field}). The contours for multiples of $2\pi$ are omitted for clarity. The picture also shows the angle $\theta$ of Cherenkov radiation. One sees that Frank's and Tamm's formula, Eq.~(\ref{angle}), gives an excellent approximation in the radiation zone characterized by condition (\ref{far}).
}
\label{phase}}
\end{center}
\end{figure}

In the far field away from the source ($z\gg r$) the vector potential is dominated by the stationary contribution with $H\sim H_0^{(1)}$. The field consists of purely outgoing radiation for positions
\begin{equation}
z\gg \frac{r}{\sqrt{(nv/c)^2-1}} 
\label{far}
\end{equation}
according to the asymptotics described by Eq.~(\ref{asymphankel}) combined with the phase factor in Eq.~(\ref{axh}). Here we obtain for the phase pattern of $\widetilde{A}_x$ the expression
\begin{equation}
\phi\sim \frac{\omega}{v}\left(z+r\sqrt{\frac{n^2v^2}{c^2}-1}\right) \,.
\label{mach}
\end{equation}
The phase fronts form cones with an angle $\theta$ with $\tan\theta = \sqrt{(nv/c)^2-1}$ such that
\begin{equation}
\cos\theta = \frac{c}{nv} \,.
\label{angle}
\end{equation}
This is Frank's and Tamm's formula for the angle of Cherenkov radiation. Figure~\ref{phase} shows the actual phase profile of $\widetilde{A}_x$ including near--field and entrance effects. One sees that the Mach cones with phase pattern of Eq.~(\ref{mach}) are an excellent approximation for the far--field regime characterized in Eq.~(\ref{far}). The numerical calculation was done after deforming the integration contour in Eq.~(\ref{h}) such that 
\begin{eqnarray}
H = H_0^{(1)}(\delta k\,r) &-& \frac{1}{\pi} \int_0^{\pi/2} e^{i\delta k r \cosh(\chi_0+i\eta)}\, d\eta
\nonumber\\
&-& \frac{1}{i\pi} \int_{\chi_0}^\infty e^{-\delta k r\sinh\chi}\, d\chi
\label{hnum}
\end{eqnarray}
that rapidly converges. The calculation of the vector potential shows that the field of the moving point polarization does indeed have the same characteristic phase pattern of Cherenkov radiation (Fig.~\ref{phase}). It remains to calculate the radiation spectrum. 

\begin{figure}[t]
\begin{center}
\includegraphics[width=20.pc]{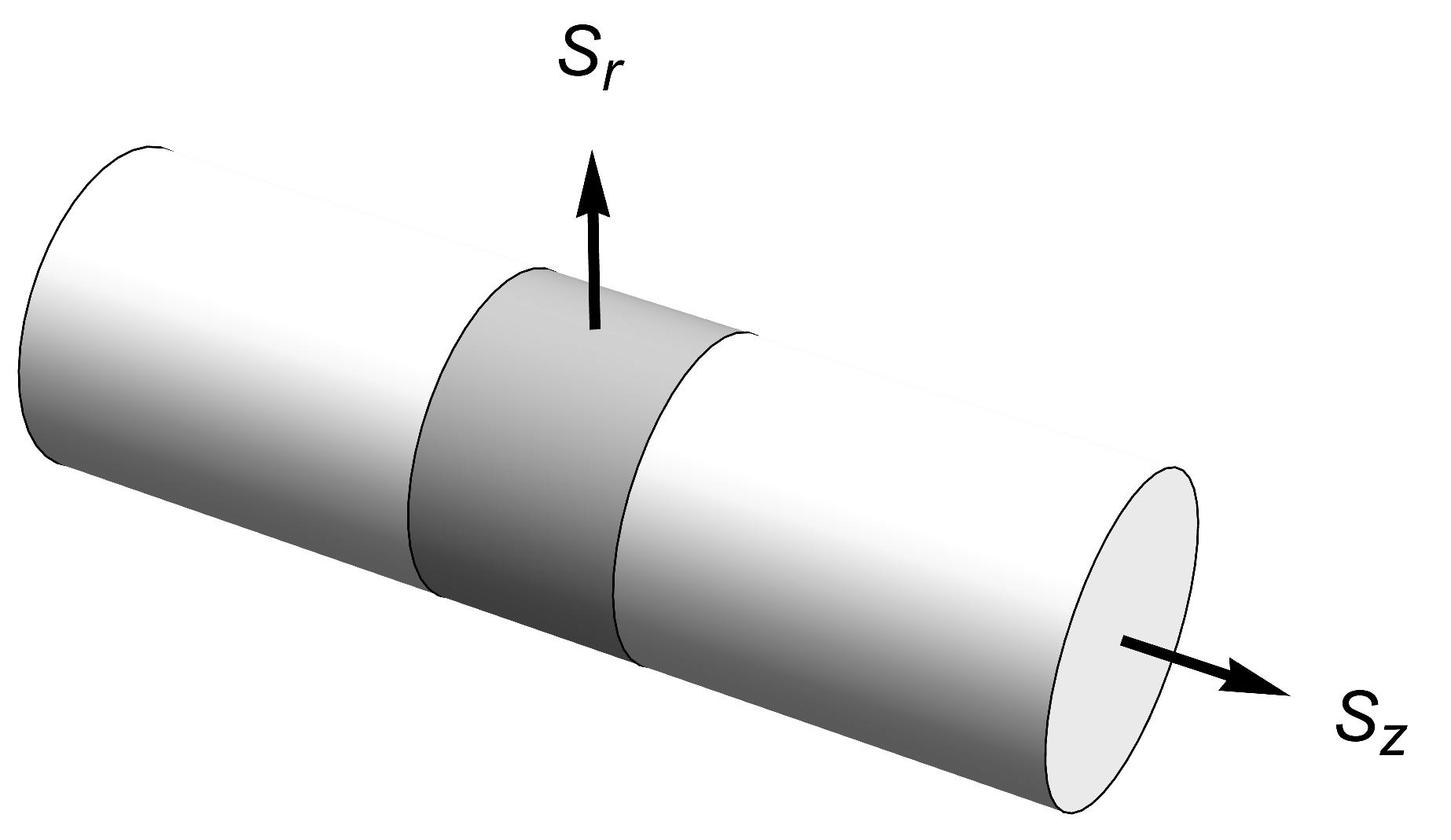}
\caption{
\small{
Cylinder of the integration surface to obtain the energy flux from the components of the Poynting vector $\bm{S}$. For getting an analytic expression of the emission spectrum the cylinder is made infinitesimally small. 
}
\label{cyl}}
\end{center}
\end{figure}

The Cherenkov spectrum is given by the energy flux across a surface around the moving polarization. For deriving an exact expression with minimal technical effort we imagine this surface as a closed cylinder around the $z$--axis (Fig.~\ref{cyl}) with radius going to zero. In this case, we need the behavior of $\widetilde{A}_x$ for $r\rightarrow 0$, {\it i.e.} in the near--field regime close to the source. We see from Eq.~(\ref{cosh}) that 
\begin{equation}
\cosh\chi_0\sim \frac{z}{r}\, \sqrt{\frac{nv-c}{nv+c}} \quad\mbox{for}\quad r\sim 0
\end{equation}
where also $\cosh\chi\sim e^\chi/2$ for which we can solve the integral in Eq.~(\ref{h}) exactly. We obtain \cite{Erdelyi}
\begin{eqnarray}
H&\sim& H_0^{(1)}(\delta k\,r) + \frac{1}{i\pi}\,\mathrm{Ei}\left(i\delta k r\,\cosh\chi_0\right)-1
\nonumber\\
&\sim& \frac{2i}{\pi}\left(\gamma +\ln \frac{\delta k\,r}{2}\right) + \frac{1}{i\pi}\,\mathrm{Ei}(i\zeta)
\label{hasymp}
\end{eqnarray}
with Euler's constant $\gamma$ and
\begin{equation}
\zeta = \frac{\omega}{c}\,z\left(n-\frac{c}{v}\right) \,.
\label{zeta}
\end{equation}
Armed with these expressions, we can calculate the emission spectrum analytically. But first we need to extract the electromagnetic field strengths from the vector potential that give the energy flux as the Pointing vector $\bm{S}$ integrated over the cylinder (Fig.~\ref{cyl}). 

\section{Field strengths}

The electromagnetic field strengths are given by Eq.~(\ref{eh}) in general, here we need them in cylindrical coordinates $\{r,\phi,z\}$. As ${\bm A}\cdot d{\bm r}=A_xdx=A_x(\cos\phi\,dr-r\sin\phi\,d\phi)$ is a spatial scalar, we read off $A_r$ and $A_\phi$ as
\begin{equation}
A_r = A_x\cos\phi \,,\quad A_\phi = -A_x r\sin\phi \,.
\end{equation}
From this and Eq.~(\ref{u}) of the potential $U$ follows
\begin{eqnarray}
\widetilde{E}_r &=& i\omega \widetilde{A}_x \cos\phi - \partial_r \widetilde{U} 
\nonumber\\
&=& \frac{i\omega}{n^2k^2} \left(\partial_r^2+n^2k^2\right) \widetilde{A}_x\cos\phi
\label{er0}
\end{eqnarray}
and from Eq.~(\ref{ax}) in cylindrical coordinates
\begin{equation}
\widetilde{E}_r = \frac{\cos\phi}{n^2}\left[\frac{c}{ik}\left(\partial_z^2+\frac{1}{r}\,\partial_r\right)\widetilde{A}_x-P\Delta\right]
\label{er}
\end{equation}
with the contact term
\begin{equation}
\Delta = \frac{\mu_0 c^2}{v}\,e^{i(\omega/v)z}\,\delta(x)\,\delta(y)\,.
\label{contact}
\end{equation}
We obtain for the other components of the electric field strength 
\begin{eqnarray}
\widetilde{E}_\phi &=& -i\omega \widetilde{A}_x r \sin\phi - \partial_\phi \widetilde{U} 
\nonumber\\
&=& \frac{\sin\phi}{n^2}\,\frac{c}{ik}\left(\frac{1}{r}\,\partial_r+n^2k^2\right)\widetilde{A}_x r \,,
\label{ephi} \\
\widetilde{E}_z &=& -\partial_z \widetilde{U} = -\frac{\cos\phi}{n^2}\,\frac{c}{ik}\,\partial_z\partial_r \widetilde{A}_x \,.
\label{ez}
\end{eqnarray}
The magnetic field strength is given by the curl of the vector potential, 
\begin{equation}
H_i = \frac{1}{\mu_0} \,g_{ij}\,\epsilon^{jkl}\,\partial_kA_l \,,
\end{equation}
written using Einstein's summation convention over repeated indices, the metric tensor $g_{ij}=\mathrm{diag}(1,r^2,1)$ of cylindrical coordinates, and the Levy--Civita tensor $\epsilon^{jkl}$ with
\begin{equation}
\epsilon^{ijk} = \frac{1}{\sqrt{g}}\,[ijk] = \frac{1}{r} \,[ijk] 
\label{levicivita}
\end{equation}
and $[ijk]$ being the complete antisymmetric symbol \cite{LP}. We thus obtain the magnetic field components:
\begin{eqnarray}
\widetilde{H}_r &=& -\frac{1}{\mu_0 r}\,\partial_z\widetilde{A}_\phi =  \frac{\sin\phi}{\mu_0}\,\partial_z\widetilde{A}_x \,,
\label{hr}\\
\widetilde{H}_\phi &=& \frac{r}{\mu_0}\,\partial_z\widetilde{A}_r =  \frac{\cos\phi}{\mu_0}\,\partial_z\widetilde{A}_x r \,,
\label{hphi}\\
\widetilde{H}_z &=& \frac{1}{\mu_0 r}\left(\partial_r\widetilde{A}_\phi-\partial_\phi \widetilde{A}_r\right) =  -\frac{\sin\phi}{\mu_0}\,\partial_r\widetilde{A}_x \,.
\label{hz}
\end{eqnarray}
Now we have everything ready for calculating the emission spectrum.

\section{Cherenkov spectrum}

The energy flux across the surface (Fig.~\ref{cyl}) is given by the time--averaged Poynting vector \cite{Jackson,LP}. One gets for the spectral energy flux, {\it i.e.} the energy flux per frequency:
\begin{equation}
S^i = \epsilon^{ijk} \,\mathrm{Re}\,\left\{\widetilde{E}_j \widetilde{H}_k^*\right\} \,.
\label{poynting}
\end{equation}
In cylindrical coordinates with Levy--Civita tensor of Eq.~(\ref{levicivita}) we have
\begin{eqnarray}
S_r &=& \frac{1}{r}\,\mathrm{Re}\,\left\{ \widetilde{E}_\phi \widetilde{H}_z^*-\widetilde{E}_z\widetilde{H}_\phi^*\right\} \,,
\label{sr} \\
S_z &=& \frac{1}{r}\,\mathrm{Re}\,\left\{ \widetilde{E}_r \widetilde{H}_\phi^*-\widetilde{E}_\phi\widetilde{H}_r^*\right\} 
\label{sz}
\end{eqnarray}
where we lowered the index without change for the $r$ and $z$ components in cylindrical coordinates. The $\phi$--component of the Poynting vector vanishes in our case, as the radiation does not cycle around the propagation axis. 

Consider the differential spectral energy $\sigma$ per propagation length:
\begin{equation}
\sigma=\frac{d^2W}{d\omega\, dz}
\end{equation}
where $W$ denotes the energy. According to Poynting's theorem \cite{Jackson} we can write $dW/d\omega$ as the surface integral of the Poynting vector $\bm{S}$ given by Eq.~(\ref{poynting}). The surface we can deform in any way, as long as it encloses the interval of the propagation (on the $z$--axis) we wish to consider, because Cherenkov radiation is produced only at the propagation axis. For a point source, it will be advantageous to employ a cylinder (Fig.~\ref{cyl}) with vanishing radius $r_0$. In this case, we have
\begin{equation}
\sigma=r_0 \int_0^{2\pi}\left. S_r\right|_{r_0} d\phi +\partial_z \int_0^{r_0}\int_0^{2\pi} S_z\,r\, d\phi\,dr
\label{sigma}
\end{equation}
for $r_0\rightarrow 0$. We obtain for the first term, {\it i.e.} for the differential flux in radial direction:
\begin{equation}
\int_0^{2\pi} S_r \,d\phi = -\frac{\pi\omega}{\mu_0} \,\mathrm{Im}\left\{ (\partial_r\widetilde{A}_x^*)\widetilde{A}_x +\frac{(\partial_r\partial_z\widetilde{A}_x^*)\partial_z\widetilde{A}_x}{n^2k^2}\right\}
\label{rflux}
\end{equation}
where we used Eqs.~(\ref{ephi}), (\ref{ez}), (\ref{hphi}) and (\ref{hz}) in Eq.~(\ref{sr}). Now we turn to the flux in propagation direction. Here only the contact term [Eq.~(\ref{contact})] in the radial component of the electric field, Eq.~(\ref{er}), can make a contribution for $r\rightarrow 0$, for the following reason. The asymptotics described in Eq.~(\ref{hasymp}) implies that the other terms are diverging logarithmically or with $1/r$ at most. It turns out that the $1/r$ terms cancel each other such that only the logarithmic divergency remains, but the integral of a logarithm over an infinitesimally small disk vanishes. In this way we obtain from Eqs.~(\ref{er}-\ref{ephi}), (\ref{hr}), (\ref{hphi}) and (\ref{sz}) the spectral energy flux in propagation direction:
\begin{equation}
\int_0^{r_0}\int_0^{2\pi} S_z\,r\, d\phi\,dr = -\frac{c^2 P}{2 n^2 v}\,\mathrm{Re}\left\{e^{i(\omega/v)z}\partial_z \widetilde{A}_x^*\right\}
\label{zflux}
\end{equation}
in the limit $r_0\rightarrow 0$. We have expressed the fluxes in terms of the vector potential. According to Eq.~(\ref{axh}) the vector potential depends on the Hankel--type amplitude $H$. We obtain from Eq.~(\ref{hasymp}) 
\begin{equation}
\partial_r H \sim \frac{2i}{\pi r} \,,\quad \partial_z H \sim \frac{1}{i\pi z}\,e^{i\zeta}
\end{equation}
with $\zeta$ given by Eq.~(\ref{zeta}), express $\mathrm{Ei}(i\zeta)$ as $\mathrm{Ci}(\zeta)+i\,\mathrm{Si}(\zeta) +i\pi/2$ according to Ref.~\cite{Erdelyi} and obtain for the emission spectrum from Eqs.~(\ref{axh}), (\ref{rflux}) and (\ref{zflux}) the exact expression 
\begin{equation}
\sigma = \frac{\mu_0 P^2\omega^3}{8v^2\pi}\left[\left(1+\frac{c^2}{n^2v^2}\right)
\left(\mathrm{Si}(\zeta)+\frac{\pi}{2}\right)
-\partial_z\frac{\sin\zeta}{n^2 k^2 z}\right] 
\label{sigmatransient}
\end{equation}
with $\zeta$ being defined in Eq.~(\ref{zeta}). The spectrum contains the transient radiation \cite{TransitionRad} due to the light bullet suddenly entering the medium at $z=0$. For large $z$ the spectrum approaches a stationary value, because \cite{Erdelyi} $\mathrm{Si}(\zeta)\sim \pi/2 -(\cos\zeta)/\zeta$ and terms falling with $\zeta^{-1}$ or $z^{-1}$ or stronger are vanishing. The stationary limit of the spectrum $\sigma$ for $z\rightarrow\infty$ describes the stationary Cherenkov radiation for $v\ge c/n$. Furthermore, we know \cite{Condition} that $\sigma$ vanishes for stationary fields at frequencies below the Cherenkov threshold ($v<c/n$). In the stationary regime we thus obtain the remarkably simple result:
\begin{equation}
\sigma = \begin{cases}
\displaystyle \frac{\mu_0 P^2\omega^3}{8v^2}\left(1+\frac{c^2}{n^2v^2}\right) & \text{for}\quad v\ge c/n \\
\displaystyle 0 & \text{otherwise}
\end{cases}
\,.
\label{result}
\end{equation}
Frank's formula, Eq.~(4.36) of Ref.~\cite{Frankreview}, for the Cherenkov spectrum of a superluminally fast magnetic dipole polarized orthogonally to the propagation direction, is more complicated, but it shares the same characteristic features with our simple result, Eq.~(\ref{result}). The emission spectrum grows with the cube of the frequency and it differs from zero already at the threshold where $v=c/n$. In contrast, the spectrum of an electric or a magnetic dipole pointing parallel to the propagation direction is  \cite{Frankreview} 
\begin{equation}
\sigma_\parallel = \begin{cases}
\displaystyle\frac{\mu_0 P^2\omega^3}{8v^2}\left(1-\frac{c^2}{n^2v^2}\right) & \text{for}\quad v\ge c/n \\
\displaystyle 0 & \text{otherwise}
\end{cases}
\label{parallel}
\end{equation}
and for an electric dipole orthogonal to the direction of motion \cite{Frankreview}
\begin{equation}
\sigma_\bot = \begin{cases}
\displaystyle \frac{\mu_0 P^2n^2\omega^3}{16c^2}\left(1-\frac{c^2}{n^2v^2}\right)^2 & \text{for}\quad v\ge c/n \\
\displaystyle 0 & \text{otherwise}
\end{cases}
\,.
\label{orthogonal}
\end{equation}
In Eqs.~(\ref{parallel}) and (\ref{orthogonal}) $P$ accounts for the dipole moment such that the formulas are adjusted to Eq.~(\ref{result}). With our theoretical calculation of the Cherenkov radiation of a point--like light bullet we have thus confirmed Frank's puzzling result \cite{Frank,Frankreview} for the combination of electric and magnetic dipole the light bullet corresponds to (as explained in Sec.~II).

\section{Experimental evidence?}

\begin{figure}[t]
\begin{center}
\includegraphics[width=20.pc]{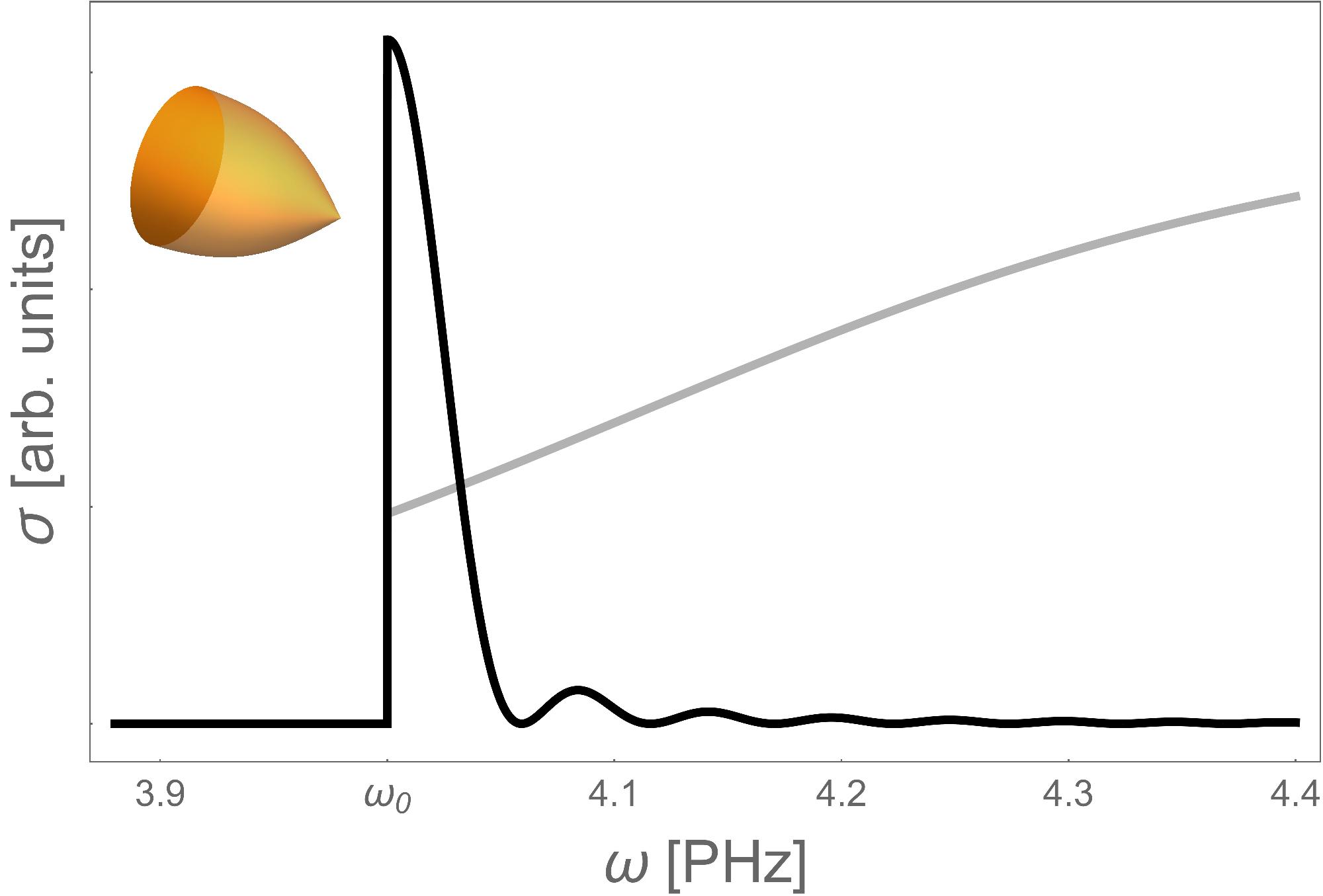}
\caption{
\small{
Peak of Cherenkov radiation from an extended light bullet (black curve) versus the spectrum of a point source (gray curve). One sees how the discontinuity at the threshold for the point object is turned into a peak for the extended source. The spectrum was plotted according to Eq.~(\ref{extendedresult}); the gray curve for the point source was obtained by integrating $\sigma$ of Eq.~(\ref{sigmatransient}) and Fig.~(\ref{spectra}c) from 0 to $z$. The parameters are $a=5\mu\mathrm{m}$, $z=1000\mu\mathrm{m}$, the phase velocity $c/n$ in glass at $\omega_0=4\mathrm{PHz}$ was taken as velocity $v$ of the moving polarization. The standard Sellmeier formula for fused silica \cite{Sellmeier} was used for $n(\omega)$.}
\label{peak}}
\end{center}
\end{figure}

In order to directly compare the Cherenkov radiation of light bullets with Frank's moving magnetic dipoles we made one idealization in our model that is currently unrealistic in practice: we assumed the light bullet to be a point object. In reality, a light bullet \cite{Silberberg} or a related optical filament \cite{Facciobook} extends over several wavelenghts. We can imagine it as a collection of point objects, but the Cherenkov radiation emitted from all these points is going to interfere and cancel each other out, unless the radiation pattern is completely frozen in the co--moving frame, which is only the case at threshold where $c/n=v$. Our result for the Cherenkov spectrum, Eq.~(\ref{result}), shows that even at threshold the emitted energy does not vanish. Therefore, the extended light bullet will still radiate, but the jump at the threshold (Fig.~\ref{spectra}c) will turn into a peak (Fig.~\ref{peak}). Such a peak has been observed in a pioneering experiment \cite{Faccio1} attempting to detect the analogue of Hawking radiation \cite{Hawking} with moving light filaments \cite{Facciobook} playing the role of the event horizon \cite{Philbin}. Our theory indicates that instead of Hawking radiation the experimentalists \cite{Faccio1} had seen the optical equivalent of Frank's elusive Cherenkov radiation of transversal magnetic dipoles.

In order to make our point more quantitative, we are going to describe the Cherenkov radiation of an extended light bullet. We will not attempt to re--create the realistic situation in the computer --- this has been partly done before \cite{FaccioSupp} --- but rather use a simple, characteristic model for working out the essential features analytically and for being more general than a specific experiment. Let us assume the moving polarization sits primarily in a planar disk corresponding to the back plane of the light bullet where the Cherenkov threshold is reached \cite{KerrComment}. For describing the effective extension of the disk we use a Gaussian multiplied with a plane of constant polarizations pointing in $x$--direction:
\begin{equation}
P_x = \frac{P}{2\pi a^2}\, \exp\left(-\frac{r^2}{2a^2}\right)\Theta(z)\,\delta(z_0) 
\label{disk}
\end{equation}
where $a$ accounts for the size of the disk. The disk is assumed to be infinitely thin and moving with $z_0=z-vt$ as the point--like source considered before. The Heaviside function models the entrance of the light bullet into the host medium of refractive index $n$. We are going to show that a disk larger then the wavelength suppresses Cherenkov radiation by destructive interference, except at threshold. In a three--dimensionally extended light bullet the emission from different planes will share the same fate. Therefore we expect that our planar model describes the essence of the Cherenkov radiation of extended light bullets.

Having established our model, we proceed to calculate the vector potential and the emission spectrum. The Fourier--transformed vector potential $\widetilde{A}_x$ is the convolution of our solution for the point source with the Gaussian of Eq.~(\ref{disk}). Using the cylindrical symmetry of our case we represent $\widetilde{A}_x$ as the spatial Bessel transform of the Fourier--transformed Gaussian with the spatial Fourier transform ${\cal A}$ of the point solution as 
\begin{equation}
\widetilde{A}_x = \frac{1}{2\pi} \int_0^\infty \exp\left(-\frac{a^2u^2}{2}\right) J_0(u r)\, {\cal A} \,u\,du
\label{uint}
\end{equation}
with Bessel function $J_0$ \cite{Erdelyi}. For the spatial Fourier transform ${\cal A}$ of the point source we employ the same expression in terms of the Green function as before, Eq.~(\ref{axg}), but replace $\widetilde{G}$ by its spatial Fourier transform $\widetilde{\cal G}$ satisfying
\begin{equation}
\left(\partial_z^2 -u^2 + n^2k^2\right) \widetilde{{\cal G}} = i\omega \delta(z) \,.
\end{equation}
This ordinary differential equation has the causal solution
\begin{equation}
\widetilde{{\cal G}} = \frac{\omega}{2\beta}\,e^{i\beta|z|}
\end{equation}
in terms of the effective wave number
\begin{equation}
\beta = \sqrt{n^2k^2-u^2} \,.
\label{beta}
\end{equation}
Solving the integral in the equivalent of Eq.~(\ref{axg}) gives the spatial Fourier transform of the vector potential of the point source, as required in Eq.~(\ref{uint}):
\begin{eqnarray}
{\cal A} &=& -\frac{i\omega \mu_0 P}{2\beta}\, f \,, \\
f &=& \frac{e^{i(\omega/v)z} - e^{i\beta z}}{\omega-v\beta} - \frac{e^{i(\omega/v)z} }{\omega+v\beta} \,.
\label{f}
\end{eqnarray}
Consider a light bullet much larger than the wavelength,
\begin{equation}
a \gg \frac{2\pi}{nk} \,.
\label{large}
\end{equation}
In this case the Gaussian in the integral of Eq.~(\ref{uint}) suppresses the values of $f(\beta,z)$ for $\beta\neq nk$ where $u\neq 0$ according to Eq.~(\ref{beta}). We thus regard
\begin{equation}
\frac{f(\beta,z)}{\beta} \sim \frac{f(nk,z)}{nk}
\end{equation}
and obtain, after performing the remaining Gaussian integral in Eq.~(\ref{uint}), the simple formula
\begin{equation}
\widetilde{A}_x \sim \frac{c\mu_0 P}{4\pi i n a^2}\, \exp\left(-\frac{r^2}{2a^2}\right) f(nk,z) \,.
\label{axtended}
\end{equation}
Now we are ready for calculating the Cherenkov spectrum of the extended light bullet. 

As the field described by Eq.~(\ref{axtended}) is concentrated in a Gaussian cylinder along the propagation direction on the $z$--axis, no radiation goes out in radial direction sufficiently far away from the $z$--axis. It is advantageous to adjust the integration surface of the energy flux. For the point source we employed an infinitely thin cylinder (Fig.~\ref{cyl}), for the extended source we now use as a convenient integration surface an infinitely thick cylinder where, as the field exponentially vanishes for $r\rightarrow\infty$ due to interference, no radiation goes out through the side. The energy flux is thus given by the difference between the flux through the top and the bottom of the infinitely thick cylinder:
\begin{equation}
\frac{dW}{d\omega} = \int_0^z \sigma\,dz = \int_0^\infty \int_0^{2\pi}\left(\left. S_z\right|_z-\left. S_z\right|_0\right) d\phi\, r dr \,.
\end{equation}
We obtain from Eqs.~(\ref{er0}), (\ref{ephi}), (\ref{hr}), (\ref{hphi}) and (\ref{sz}):
\begin{equation}
\int_0^{2\pi} S_z \, d\phi = -\frac{\pi\omega n^2}{\mu_0}\,\mathrm{Im}\left\{(\partial_z\widetilde{A}_x^*)\left(2+\frac{\Delta^{(2)}}{n^2k^2}\right)\widetilde{A}_x\right\}
\end{equation}
in terms of the 2D--Laplacian
\begin{equation}
\Delta^{(2)}= \partial_r^2+\frac{1}{r}\partial_r \,.
\end{equation}
In the limit of a large light bullet the Laplacian is significantly smaller than $n^2k^2$ such that we can safely ignore the term. Hence we get for the total flux:
\begin{eqnarray}
\int_0^\infty \int_0^{2\pi} &&\!\!\!\!\!\!\!\! S_z \, d\phi\, r dr \sim \frac{c^2\omega \mu_0 P^2}{16\pi a^2}\, \mathrm{Im}\left\{(\partial_z f^*) f\right\}
\nonumber\\
&=& \frac{c^3 \mu_0 n P^2}{4\pi a^2}\,\frac{\sin^2(\zeta/2)-\frac{(c-nv)^2}{4(c+nv)^2}}{(c-nv)^2} \end{eqnarray}
with $\zeta$ defined in Eq.~(\ref{zeta}). Finally, we obtain for the Cherenkov spectrum of the extended light bullet:
\begin{equation}
\frac{dW}{d\omega} = \int_0^z \sigma \,dz \sim \frac{\mu_0 n^3 P^2 c\, (kz)^2}{16\pi a^2}\, \mathrm{sinc}^2(\zeta/2)
\label{extendedresult}
\end{equation}
with $\zeta$ given by Eq.~(\ref{zeta}). Clearly, the discontinuity at the threshold of Cherenkov radiation has manifested itself as a peak growing with growing propagation distance $z$ (Fig.~\ref{peak}). 

Our theory does describe the main feature observed in the spectrum of the experiment \cite{Faccio1}, but it does not account for the fact that at least some part of the radiation reached the detector that was placed orthogonal to the propagation direction \cite{Faccio1}. Radiation emitted in other directions was not measured. Presumably, the curvature of the light bullet did bend some Cherenkov radiation sidewards where it could be detected. The moving filament \cite{Facciobook} in the experiment \cite{Faccio1} created a gradual perturbation of the refractive index (similar to a graded--index lens). Such a gradual perturbation may scatter a small fraction of light from propagation direction to the orthogonal direction of detection \cite{Faccio1}. 

In order to get a rough estimate of the order of magnitude of the Cherenkov radiation, we note that the nonlinear polarization $\bm{P}$ is given by $\varepsilon_0 \chi_\mathrm{NL} \bm{E}$ where $\varepsilon_0$ is the permittivity of the vacuum and $\chi_\mathrm{NL}\approx 2n\delta n$ the nonlinear contribution to the susceptibility that produces the change $\delta n$ of the refractive index; $\delta n \approx10^{-3}$ in  the experiment \cite{Faccio1}. The total polarization $P$ we estimate as $|\bm{P}|V$ with the effective volume $V\approx a^2\lambda$, assuming that the radiating disk has roughly the thickness of the wavelength $\lambda$. From this we get that $\mu P^2 \approx 4n^2\delta n^2 (a^2/c) T^2/T_\mathrm{p}$ times the total energy of the pulse where $T$ denotes the time of an optical cycle and $T_\mathrm{p}$ the pulse duration, $T\approx 3\times10^{-15}$ and $T_\mathrm{p}=1\mathrm{ps}$ in Ref.~\cite{Faccio1}. From Ref.~\cite{Faccio1} we also read off a spectral width of about $0.1\, T^{-1}$ that via Eq.~(\ref{extendedresult}) and the Sellmeier formula of the refractive index \cite{Sellmeier} corresponds to $kz\approx 10^3$. As the moving filament \cite{Facciobook} in the experiment \cite{Faccio1} is transient we use this method to estimate the effective propagation length $z$. With these estimations, we find that the total energy produced by Cherenkov radiation lies in the order of $\delta n^2$ of the total pulse energy, which is a sufficiently large number such that a small fraction bent sideways may have produced the observed peak at the Cherenkov frequency \cite{Faccio1}. Our estimation should be taken with a grain of salt though, because it is based on a rather simple model that is only meant to explain how a step in the Cherenkov spectrum for a point source can produce, by interference, a peak for an extended source. 

\section{Summary}

Nonlinear electric polarizations, moving faster than the speed of light in dielectric media, radiate (Figs.~\ref{field} and \ref{phase}). This radiation resembles the hitherto unobserved Cherenkov radiation of transversal magnetic dipoles Frank had been puzzled with for decades \cite{Frankreview}. While ordinary Cherenkov radiation gradually rises when charged particles, electric dipoles or parallel magnetic dipoles exceed the speed of light in the medium (Figs.~\ref{spectra}a and \ref{spectra}b), the radiation of transversal magnetic dipoles suddenly comes into being. We reproduce this sudden onset of Cherenkov radiation  (Fig.~\ref{spectra}c) for point polarizations and show that for extended sources the optical interference of the emitted radiation turns the discontinuity at the threshold into a peak (Fig.~\ref{peak}). Our study indicates that this peak was probably observed in the first attempt \cite{Faccio1} to measure Hawking radiation in an optical analogue \cite{Philbin}. It seems that instead of Hawking radiation, Frank's elusive magnetic Cherenkov radiation was seen. 

\section*{Acknowledgements}

This paper is dedicated to the memory of Yaron Silberberg. We are most grateful to him and to
David Bermudez,
Yehonathan Drori
and Itay Griniasty for scientific discussions related to this paper. 
Our work was supported by the European Research Council, the Israel Science Foundation, and the Murray B. Koffler Professorial Chair.
Both authors contributed equally to this paper.



\begin{thebibliography}{99}
\bibitem{Condition}
We recall the argument \cite{LL8} for the threshold of stationary Cherenkov radiation: For a field moving as $F(z-vt)$ with the moving source the frequency $\omega$ and spatial wavenumber $k_z$ are related by $\omega=k_z v$. Since $k_z\le k = n(\omega/c)$ we get $v\ge c/n$ as condition. Note that this argument does not apply to non--stationary fields generated by the source being accelerated or switched on, where some transition radiation is also made for frequencies with $v<c/n$ \cite{TransitionRad}.
\bibitem{KerrComment}
The Kerr effect adds a contribution to the refractive index that is proportional to the intensity of the light pulse \cite{NLO}. Due to this extra contribution the Cherenkov threshold is reached in a surface on the pulse; the strongest effect due to the most rapid change happens at the back face of the light bullet.
\end{thebibliography}
\end{document}